\documentclass[10pt,twocolumn,letterpaper]{article}

\usepackage{cvpr}              % To produce the CAMERA-READY version

\usepackage{booktabs}       % professional-quality tables
\usepackage{amsfonts}       % blackboard math symbols
\usepackage{nicefrac}       % compact symbols for 1/2, etc.
\usepackage{microtype}      % microtypography
\usepackage[table]{xcolor}
\usepackage{xcolor}         % colors
\usepackage{graphicx}
\usepackage{amsmath}
\usepackage{amssymb}
\usepackage{amsthm}
\usepackage{mathrsfs}
\usepackage{algorithm}
\usepackage{algpseudocode}
\usepackage{verbatim}
\usepackage{cite}
\usepackage{float}
\usepackage{multirow}
\usepackage{makecell}
\definecolor{cvprblue}{rgb}{0.21,0.49,0.74}
\usepackage[pagebackref,breaklinks,colorlinks,citecolor=cvprblue]{hyperref}

\def\paperID{7} % *** Enter the Paper ID here
\def\confName{CVPR}
\def\confYear{2024}

\title{Vim4Path: Self-Supervised Vision Mamba for Histopathology Images}

\author{Ali Nasiri-Sarvi$^{1}$~~~~Vincent Quoc-Huy Trinh$^{2}$~~~~Hassan Rivaz$^3$~~~~Mahdi S. Hosseini$^1$\\
$^1$Department of Computer Science and Software Engineering (CSSE), Concordia University, Canada\\
$^2$Institute for Research in Immunology and Cancer of the University of Montreal, Canada\\
$^3$Department of Electrical and Computer Engineering (ECE), Concordia University, Canada\\
}

\begin{document}
\maketitle
\begin{abstract}
Representation learning from Gigapixel Whole Slide Images (WSI) poses a significant challenge in computational pathology due to the complicated nature of tissue structures and the scarcity of labeled data. Multi-instance learning methods have addressed this challenge, leveraging image patches to classify slides utilizing pretrained models using Self-Supervised Learning (SSL) approaches. The performance of both SSL and MIL methods relies on the architecture of the feature encoder. This paper proposes leveraging the Vision Mamba (Vim) architecture, inspired by state space models, within the DINO framework for representation learning in computational pathology. We evaluate the performance of Vim against Vision Transformers (ViT) on the Camelyon16 dataset for both patch-level and slide-level classification. Our findings highlight Vim's enhanced performance compared to ViT, particularly at smaller scales, where Vim achieves an 8.21 increase in ROC AUC for models of similar size. An explainability analysis further highlights Vim's capabilities, which reveals that Vim uniquely emulates the pathologist workflow—unlike ViT. This alignment with human expert analysis highlights Vim's potential in practical diagnostic settings and contributes significantly to developing effective representation-learning algorithms in computational pathology. We release the codes and pretrained weights at \color{purple}\url{https://github.com/AtlasAnalyticsLab/Vim4Path}. 
\end{abstract}    
\section{Introduction}
\label{sec:intro}
Representation learning from Gigapixel Whole Slide Image (WSI) is a challenging problem in computational pathology. Each tissue WSI comprises hundreds of thousands of cells with biological importance, interacting with each other as neighborhoods, with microscopic juxtaposition of different cell types. To represent the holistic nature of each WSI, one needs to collect thousands of manageable instances (known as image patches) from each slide to process and interpret the tissue cells in WSI. However, the diagnostic relevance of the tissue cells within each image patch is unidentified, where only a weakly-supervised label is available on the slide-level for representation. To bridge the gap, Multi-Instance Learning (MIL) methods \cite{MIL} are adopted to aggregate embedded features from image patches and classify them on the bag level (i.e., slide) \cite{hosseini2024computational}. Most MIL methods rely on utilizing a feature extractor, such as ResNet50 \cite{ResNet50}, as a backbone model to assign an embedding to each item in the bag. Pretrained models from natural images, such as ImageNet \cite{imagenet}, were traditionally adopted for feature extraction. However, the distribution of pathology images differs from natural images, causing sub-optimal performance when using pretrained ImageNet weights. 

To address these issues, better approaches exist beyond simply fine-tuning ImageNet weights, as WSIs often use weak supervision with only one label per slide, and typically, just a few image patches contain diagnostic information. Although more granular annotations can be used (e.g., region-of-interest (ROI) or patch-level), they are costly and scarce due to the extensive expert analysis they require \cite{hosseini2024computational}. As a result, recent advancements in computational pathology have increasingly concentrated on Self-Supervised Learning (SSL) approaches \cite{kang2023benchmarking}. These methods train on extensive volumes of unlabeled data to achieve robust feature representation without relying on labels. The features obtained can then be downstream to tasks like classification, utilizing a much smaller set of labeled data. Additionally, it is important to recognize that the choice of the encoder architecture significantly influences the effectiveness of the embeddings produced by SSL methods.

Recently, a new architecture design inspired by state space models, known as Vision Mamba (Vim) \cite{Vim}, has emerged. In Section \ref{sec:methodology}, we will discuss why this architecture holds promise for pathology applications. It combines the advantageous inductive biases of CNNs, requiring less scale and computational power, with the ability to perceive long-range dependencies, similar to ViTs.  Our research intends to explore and adapt the Vim architecture within SSL schemes. To evaluate its performance, we employ DINO \cite{DINO}, a well-known SSL framework used for ViT models, as a benchmark to contrast the capabilities of Vim against ViT architectures. Our study is divided into four main objectives and findings:
\begin{enumerate}
    \item We compare the two architectures in both patch-level and slide-level classification tasks on the Camelyon16 \cite{Cam16} dataset for benchmarking.
    \item We show that the Vim model significantly outperforms the ViT model at a smaller scale.
    \item We demonstrate that the Vim model is competitive and often surpasses the ViT model's performance at scale.
    \item We perform explainability testing and discover that the Vim model mimics the pathologist's method of diagnosing different image parts.
\end{enumerate}
The rest of paper reads as follows. Section \ref{sec:related_work} discusses the related works. Section \ref{sec:preliminaries} discusses preliminaries from state space modeling. In Section \ref{sec:methodology}, we explain our proposed method and the intuition behind Vim models for pathology datasets. Finally, in Section \ref{sec:experiments}, we empirically compare the Vim and ViT architectures, assess their performance at different scales, and perform explainability analysis.

\section{Related Work}
\label{sec:related_work}
Whole-slide images (WSIs) in pathology datasets are Gigapixel images, with individual samples reaching pixel sizes up to 150,000x150,000, hindering the direct application of deep learning techniques without adequate adjustments. To manage these Gigapixel images, they are broken into smaller patches, with limited field-of-view, for processing using Multi-Instance Learning (MIL) methods \cite{MIL} such as \cite{AttMIL}, \cite{CLAM}, \cite{AB-MIL}, \cite{TransMIL}, \cite{DSMIL}, \cite{DtfdMIL}. They mainly use a pretrained encoder from natural image datasets such as ImageNet \cite{imagenet} or Pathology datasets such as Camelyon16 \cite{Cam16}. These pretrained networks on pathology datasets are usually trained in a self-supervised manner due to a lack of granular labels. The main idea behind self-supervised learning is to train an encoder without labels on a large data cohort to obtain a good feature representation, then transfer the weights for downstream tasking to train a classifier head on a smaller labeled dataset. Methods such as DINO \cite{DINO}, SimCLR \cite{SimCLR, SimCLRv2}, MoCo \cite{MoCo, MoCoV2, MoCoV3}, and MAE \cite{MAE} are some of the self-supervised learning frameworks that were introduced for natural image processing. In computational pathology, contrastive learning is used to train self-supervised models such as \cite{Dual}, \cite{s5cl}, \cite{self_histo}. HIPT \cite{hipt} uses a hierarchical structure of ViT \cite{ViT} models with each level in the hierarchy trained through DINO \cite{DINO} and ViT models \cite{ViT} architecture. A comprehensive review is provided in \cite{hosseini2024computational} on the utility of self-supervised learning in computational pathology applications. 

Recently, an alternative for ViT \cite{ViT} has been proposed based on state space models. Vision Mamba (Vim) \cite{Vim} extends state space models \cite{fu2022hungry, gu2020hippo,gu2021efficiently, gu2022parameterization, poli2023hyena, mamba} to images, delivering state-of-the-art performance with linear memory requirements. While there have been previous works in using state space models in pathology images such as \cite{ssm_pathology}, \cite{mammil}, and \cite{mambamil}, the utility of the state space models as an image encoder within the realm of computational pathology has not been yet explored.

\section{Preliminaries} \label{sec:preliminaries}
In this section, we review some concepts regarding the state space models, Vision Mamba, and Self-Supervised Learning Methods. These preliminaries are needed to discuss our proposed method in Sec \ref{sec:methodology}.

\subsection{State Space Models}
{
State space models are mathematical models for describing continuous linear systems. The state equation 
\begin{equation}
    h^\prime(t) = Ah(t) + Bx(t)
    \label{eq:state_equation}
\end{equation}
captures how the state $h(t)$ changes overtime based on the input $x(t)$. The output equation 
\begin{equation}
    y(t) = Ch(t)+Dx(t)
    \label{eq:output_equation}
\end{equation}

links the output to the hidden state and input. 

This model based on $A$, and $B$ is for continuous input. However, in deep learning models, the inputs are discrete. To discretize the model, a time scale parameter $\Delta$ is used to get discrete parameters $\bar{A}$ and $\bar{B}$ followed by
\begin{equation}
    \begin{aligned}
        &\bar{A} = exp(\Delta A), \quad \bar{B} = (\Delta A)^{-1} (exp(\Delta A)-I) . \Delta B 
    \end{aligned}
    \label{eq:discrete}
\end{equation}

In the S4 model \cite{S4}, the parameters $A, B, C, \text{and } \Delta$ are time-invariant, meaning they are defined independently of each individual token in the sequence. This design is mainly to prevent slow down of recurrent modeling, allowing this method to use a global convolution to model the sequential data. While this time-invariance helps with processing speed, it prevents the model from behaving based on the input, hence limiting its performance.  

% In structured state space models (S4) \cite{S4}, to prevent from slow down of recurrent modelling, the authors use a global convolution to model this sequential data (omitting D parameter):

% It is possible to use this global convolution since the parameters $A, B, C, \text{and } \Delta$ are time-invariant, meaning they are defined independent of each individual token in the sequence. 

% \subsection{Mamba}
% In the selective state space (Mamba) \cite{Mamba}, the parameters $B, C, \text{and } \Delta$ are computed based on the input sequence using a linear projection, making them time-variant, leaving $A$ the only time-invariant parameter. This prevents the usage of a global convolution, hence forcing this method to use recurrent processing which does not utilize the parallelization of GPUs. 

% To increase the processing speed, this method uses a hardware-aware algorithm considering that modern GPUs have to types of memory, SRAM and HBM with the first one being faster but having less capacity. Since the parameter $A$ is still time-invariant, they move it the the fast SRAM, for the sequential processing, at each time step, they compute $B, C, \text{and } \Delta$ in HBM, move $B \text{ and } \Delta$ to SRAM to compute Eq. \ref{eq:discrete} and the discrete version of Eq. \ref{eq:state_equation}. Following this, the state vectors are moved to HBM to compute the Eq. \ref{eq:output_equation} using the previously computed C (D is omitted). This hardware-aware algorithm makes Mamba's throughput comparable to the parallel transformers. 
}

\begin{figure*}[htp]
    \centering
    \includegraphics[width=1\textwidth]{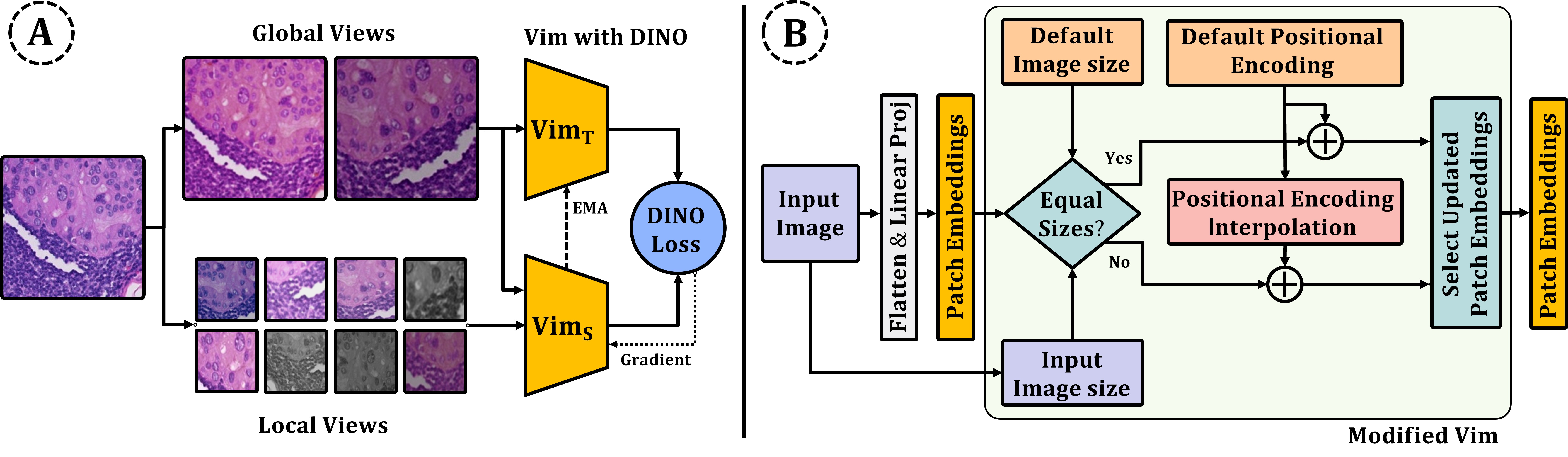}
    \caption{Detailed architecture of VIM within the DINO framework. We modify the Vim model to adapt to input image size for positional embedding interpolation and employ the modified model within DINO as a backbone architecture for self-supervised learning.}
    \label{fig:VIM4Path}
\end{figure*}
\subsection{Vision Mamba}
{
In the selective state space (Mamba) \cite{mamba}, the parameters $B, C,$ and $\Delta$ are computed based on the input sequence using a linear projection, making them time-variant, leaving $A$ the only time-invariant parameter. This prevents the usage of a global convolution, forcing this method to use recurrent processing, which does not utilize the parallelization of GPUs. To increase the processing speed, this method uses a hardware-aware algorithm to utilize GPUs for recurrent modeling. Mamba \cite{mamba} is developed for 1D sequences such as text and audio. Vision Mamba (Vim) \cite{Vim} extends this method to images by extracting small image tiles (for example, $16\times16$ tiles, which are also known as tokens), projecting them using a linear layer to an embedding dimension similar to \cite{ViT}. Furthermore, to incorporate positional awareness, they add positional encoding to the extracted projections similar to \cite{ViT}. Since Mamba \cite{mamba} is a sequential model, it uses a bidirectional Mamba to better capture the representation for a given token based on future and previous tokens since we have access to all tokens in images. 
}

\subsection{Self-Supervised Learning using DINO}
{
DINO \cite{DINO} utilizes self-distillation in a teacher-student setup, where both networks share the same architecture but differ in parameters. The teacher's outputs guide the student's training without labeled data. Training involves cropping the input image into global and local views and applying augmentations to each view, similar to SimCLR. The student predicts the teacher's output from all views, while the teacher only sees the global ones. Teacher's parameters are updated as an exponential moving average of the student weights, ensuring stability and enhancing representation quality. This method's emphasis on global-local view training is crucial for pathology images, as it helps in understanding the ``global to local'' relationships vital for detecting transitions in cancer states across neighboring cells, offering a more accurate representation of disease progression.
}

\subsection{Multi-Instance Learning using CLAM}
We use CLAM \cite{CLAM} as a framework to compare different architectures on slide-level classification. This method employs a data-efficient approach under a weakly supervised setting. It utilizes attention-based multiple-instance learning, allowing it to identify sub-regions within the slides most indicative of the slide-level label, effectively focusing on the most informative features without requiring detailed annotation. Our reason for using it is its ability to achieve high diagnostic accuracy with minimal data and labeling effort.

\section{Methodology}\label{sec:methodology}
This section elaborates on our proposed method for adopting a modified Vim \cite{Vim} with Positional Encoding Interpolation \cite{ViT} within the DINO \cite{DINO} framework. We also explain why Vim \cite{Vim} is a good alternative to ViT \cite{ViT} architecture. 
\label{sec:methodology}

\subsection{Vision Mamba within DINO}
{
Since the DINO framework requires different image sizes (e.g., $224\times224$ and $96\times96$), any encoder in this framework should be able to receive different image sizes. This is not the case with the Vanilla Vim \cite{Vim}, and we introduce a Positional Encoding Interpolation module similar to \cite{ViT} to the Vim \cite{Vim} model. The positional encoding interpolation module works by dynamically resizing the positional encoding to match the dimensions of the input image, ensuring spatial information is accurately maintained across different scales. This is achieved by using bicubic interpolation to adjust the patch embeddings according to the calculated target grid size derived from the input dimensions and patch size. This adaptation allows the default Vim model to handle images of varying sizes, allowing us to adopt it within the DINO \cite{DINO} framework. Detailed architecture of Vim within the DINO framework using the Positional Encoding Interpolation is shown in Fig \ref{fig:VIM4Path}.
}

\subsection{Architecture Design Comparison}
{
CNNs are designed with an inductive bias that processes adjacent areas within an image, such as segments A and B (Fig \ref{fig:arch_compare}), using fixed-size kernels. However, these models struggle with long-range dependencies due to their limited receptive field. This constrains their capacity to directly relate distant areas (e.g. segment A with Z) depicted in Fig \ref{fig:arch_compare}. In contrary, ViTs lack this inductive bias, treating every image patch the same, with only positional encoding differing to retain spatial relationships. This design allows ViTs to effectively capture long-range dependencies, such as the connection between segments A and Z in Fig \ref{fig:arch_compare}. However, this comes at the cost of requiring more data and computational resources to learn the nuances of local features, which are naturally handled by the inductive bias in CNNs.

Vim combines the strengths of both CNNs and ViTs. By employing sequential processing, Vim uses an inductive bias focusing on local dependencies between segments like A and B in Fig \ref{fig:arch_compare}. Additionally, Vim's usage of Mamba \cite{mamba} overcomes the short context limitation inherent in previous image sequential models like PixelRNN \cite{pixelrnn}, making it capable of capturing long-range dependencies across images, similar to ViT's capacity to relate segment A with segment Z in Fig \ref{fig:arch_compare}. This combination allows Vim to leverage the benefits of both localized and global information in images, as it sequentially scans through the image data while maintaining a long-range perspective, as demonstrated in Fig \ref{fig:arch_compare}. 

\begin{figure}[htp]
  \centering
  \includegraphics[width=1.0\linewidth]{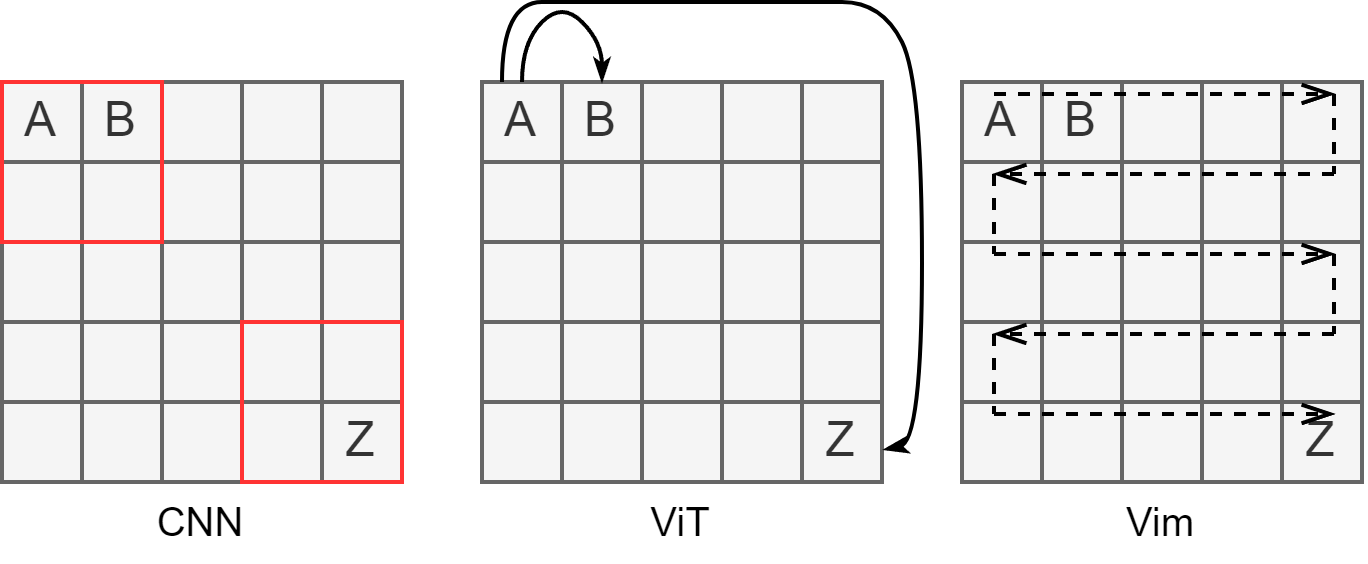}
   \caption{Comparison between different architecture designs. Vim sequential processing allows the model to capture both short-range and long-range dependencies.}
   \label{fig:arch_compare}
\end{figure}

\subsection{Vim Relationship to Pathology}
Sequential processing of Vision Mamba on patch level from the whole slide image is depicted in Fig \ref{fig:seq}. Each patch is divided into multiple tokens, and the Vision Mamba raster scans the tokens from each patch in sequence form following a lawnmower pattern. This scan enables the encoder to pick up short-range dependencies for representation in patch embedding. Later, the embedded features are aggregated on the slide level for MIL representation. This process imitates the realistic representation of cancer cells in sequence, from local zooming to the global representation on the slide, where a similar slide navigation is used by the pathologists for diagnosing cancer cells under the microscopy \cite{molin2015slide}.

\textbf{Pathology insights to short-range dependencies.} Capturing both short- and long-range dependencies in pathology images is crucial. In fact, adding the sequences for short-range dependencies at higher magnification (e.g. 10x) is more important compared to low magnification (e.g. 5x). The most important event is the sequence of cells next to each other at higher magnification, as that is what a pathologist actually practice for clinical diagnosis. We empirically demonstrate this effect in our experiments in Section \ref{sec:experiments}. While sequences at lower magnification are relevant, at high magnification it is where it becomes essential. To elaborate this in detail, during cancer evolution, normal cells throughout the body incur an initial driver mutation into cancerous cells \cite{quail2013microenvironmental}. This transition does not occur in isolation within a single cell population as coordinated responses from neighboring cells shape the tissue into sequential patterns that interact with the mutation-burdened tumor cells \cite{quail2013microenvironmental}. These sequences of neighboring cells are at the core of modern clinical pathology and genomics research \cite{binnewies2018understanding}. For instance, the sequence of lymphocytes adjacent to tumor cells and their expression of the protein PD-L1 underlie response to immune checkpoint inhibitors, a discovery based on the proximity of these immune cells to cancer, at the basis of the 2018 Nobel Prize in Physiology or Medicine \cite{waldman2020guide}. Indeed, the most novel and costly form of genomics research, termed ``spatial transcriptomics,'' was developed to overcome the lack of tools that interrogate cell networks and cell-to-cell sequences on pathology tissues \cite{moses2022museum}. In a very short period since its inception, adding these spatial sequences has led to thousands of novel discoveries \cite{moses2022museum}. Overall, there is a medical and biological need for computational pathology to deploy algorithms that query neighboring events. While both ViT and Vim are capable of encoding long-range dependencies, applying short-range dependencies is key to unraveling the complex cell networks in pathology and medicine.

\begin{figure}[htp]
  \centering
  \includegraphics[width=1.0\linewidth]{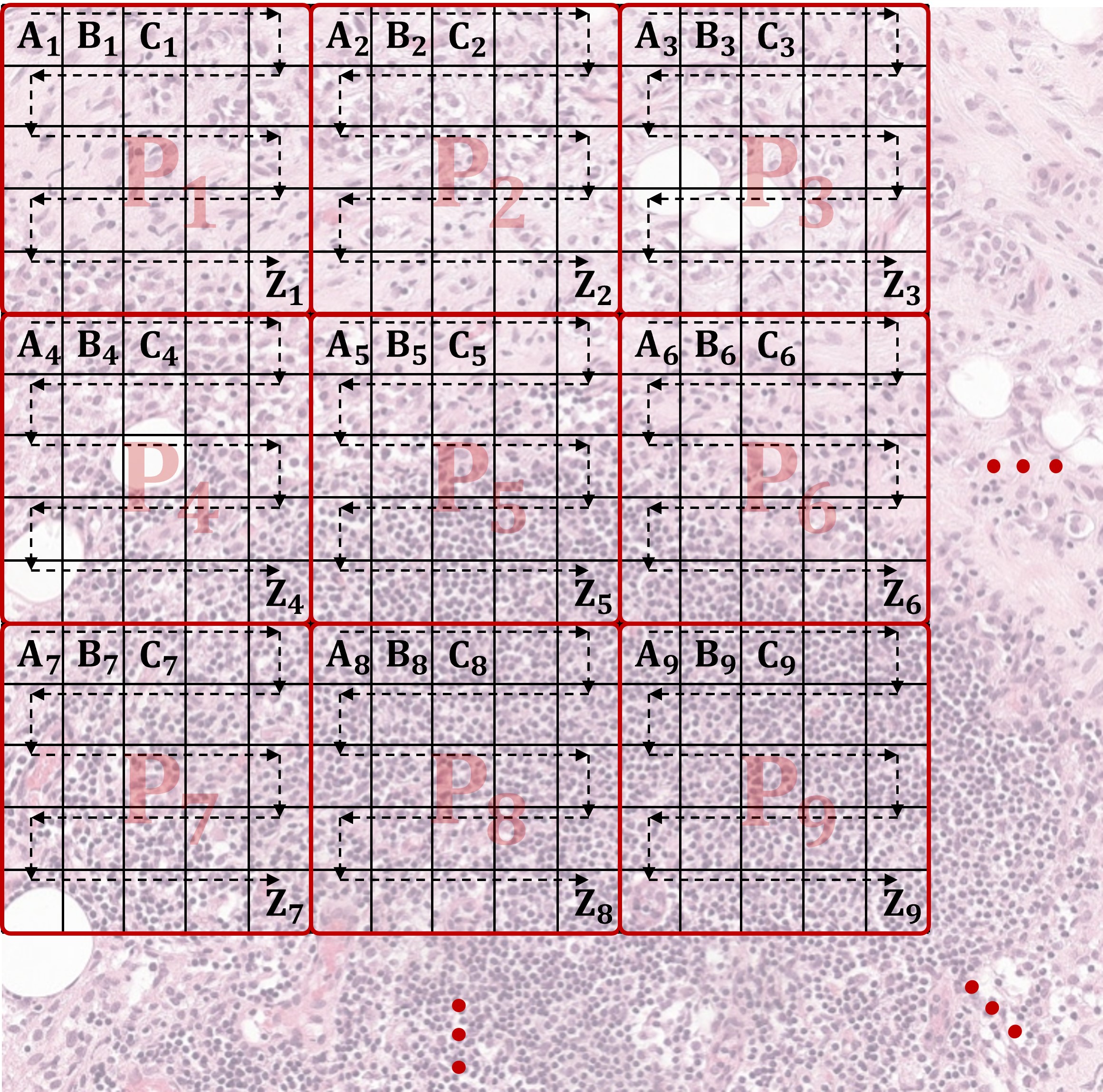}
   \caption{Sequential processing of Vim done on each patch level from slide for feature embedding. This is similar to the lawnmower  pattern used for slide navigation by pathologists to study cellular neighbourhoods in the tissue for cancer diagnosis. The information from each patch (i.e. embeddings) are put together to reach to a consensus on the slide level (i.e. aggregation).}\vspace{-.2in}
   \label{fig:seq}
\end{figure}

}
\section{Experiments}\label{sec:experiments}
In this section, we explain in detail the procedures and results of our experiments. Specifically, we outline the dataset used and the processing pipeline for this dataset, our experimental setup, and the results obtained from our experiments.

\subsection{Dataset}
We used Camelyon16 \cite{Cam16} dataset as a benchmark for our study, as extensively described in \cite{Cam16}. Table \ref{tab:cam16_distribution} summarizes the detailed distribution across different categories for the used dataset.

\begin{table}[htp]
\centering
\begin{tabular}{l|c|c}
\Xhline{1.0pt}
Category    & Training Set & Testing Set \\ \Xhline{1.0pt}
Normal Slides & 158           & 80           \\ \hline
Tumor Slides  & 110           & 49           \\ \Xhline{1.0pt}
\end{tabular}
\caption{Distribution of the used Camelyon16 dataset}
\label{tab:cam16_distribution}
\end{table}

For the pre-training phase, patches are only extracted from the training set slides, with label information being ignored, using the pre-processing pipeline of CLAM \cite{CLAM} with the default hyper-parameters. Table \ref{tab:patch_distribution} summarizes the distribution of the extracted patches for pre-training across two zooming levels. Note that the tumor patches are obtained from tumor slides without reference to ROI (region of interest) labels, indicating that not all extracted images necessarily represent tumor patches. The main reason for using 5x and 10x zooming levels was to process the data with limited computation power. We aim to compare ViT and Vim models in the context of pathology images using the same experimental setup. Hence, the 20x zooming level, which is the default for many pathology methods, was not explored in this study (due to computation limits).

\begin{table}[htp]
\centering
\begin{tabular}{l|c|c}
\Xhline{1.0pt}
Zooming Level    & Normal Patches & Tumor Patches \\ \Xhline{1.0pt}
5x &  146,508          & 21,446           \\ \hline
10x  & 555,231           & 446,608           \\ \Xhline{1.0pt}
\end{tabular}
\caption{Distribution of the pretraining dataset used. Note that the tumor patches are extracted from tumor slides without considering ROI (region of interest) labels, hence each individual image is likely not a tumor patch.}
\label{tab:patch_distribution}
\end{table}

To test each model's patch classification performance, we perform similarly to the PCam dataset \cite{Pcam} to ensure we have a balanced patch classification dataset with labels. We do not use the original PCam \cite{Pcam} dataset since the image sizes are $96 \times 96$. For the normal slides, we extract patches as before, using the pre-processing pipeline of CLAM \cite{CLAM}. We extract patches from ROI regions with $10\%$  overlap for the tumor slides. We ensure that the extracted patches have a minimum of $50\%$ intersection with the ROI region. Finally, we select a subset of patches from the normal patches to ensure a balanced dataset by uniformly sampling from all the slides to ensure maximum slide diversity among the chosen patches. We further validate the labels of the extracted patches through an expert pathologist. For the $10$x zooming level with a patch image size of $224$, we call this dataset PCam-224-10x. The statistics of these datasets are available in Table \ref{tab:pcam_distribution}.

\begin{table}[htp]
\centering
\begin{tabular}{l|c|c}
\Xhline{1.0pt}
Dataset    & Training Patches & Testing Patches \\ \Xhline{1.0pt}
PCam-224-5x  & $2\times7602$ & $2\times6611$  \\ \hline
PCam-224-10x  & $2\times28955$ & $2\times25370$ \\ \Xhline{1.0pt}
\end{tabular}
\caption{Distribution of the extracted patch classification datasets. We ensure to have the same number of patches in both classes (shown by $2\times$) }
\label{tab:pcam_distribution}
\end{table}

\begin{table}[htp]
    \centering
    \begin{tabular}{l|c}
        \Xhline{1.0pt} 
        Model &  \begin{tabular}{@{}c@{}}Parameters   (Millions) \end{tabular}  \\ \Xhline{1.0pt}
        ViT-ti & 6  \\ \hline
        ViT-s &  22 \\ \hline
        \rowcolor{green!25}Vim-ti &  7 \\ \hline
        \rowcolor{green!25}Vim-ti-plus &  13 \\ \hline
        \rowcolor{green!25}Vim-s &  26 \\ \Xhline{1.0pt}
    \end{tabular}
    \caption{Number of Trainable Parameters for each model (rounded to the nearest Million)}
    \label{tab:model_parameters}
\end{table}

\begin{table*}[htp]
\centering
\begin{tabular}{l|c|l|l|c|c|c|c|c}
\Xhline{1.0pt}
\begin{tabular} {@{}c@{}} Aggregator\\Model \end{tabular}  & \begin{tabular}{@{}c@{}}Zooming \\  Level \end{tabular} & Encoder & \begin{tabular} {@{}c@{}} Pretrained\\Dataset \end{tabular}& \begin{tabular} {@{}c@{}} Pre-training\\Method \end{tabular}& \begin{tabular} {@{}c@{}} \# Params\\   (Millions) \end{tabular}  &  ACC  & F1-Score & AUC \\ 
\Xhline{1.0pt}
Max-Pooling &20x  & ResNet50& ImageNet& Supervised& 25 & 78.95 & 71.06 & 81.28  \\ 
Mean-pooling & 20x & ResNet50& ImageNet& Supervised& 25 &76.69 & 70.41& 80.07  \\
AB-MIL \cite{AB-MIL}& 20x& ResNet50& ImageNet& Supervised& 25 & 90.06  & 87.40 & 94.00\\ 
DSMIL \cite{DSMIL} & 20x& ResNet50& ImageNet& Supervised& 25 &90.17  &  87.65 & 94.57\\
CLAM-SB \cite{CLAM} & 20x& ResNet50& ImageNet& Supervised& 25 &90.31 & 87.89 & 94.65 \\ 
CLAM-MB \cite{CLAM} &  20x& ResNet50& ImageNet& Supervised& 25 &90.14 &  88.10 & 94.70 \\ 
TransMIL \cite{TransMIL} & 20x& ResNet50& ImageNet& Supervised& 25 & 89.22 &  85.10  & 93.51 \\ 
DTFD-MIL \cite{DtfdMIL} & 20x& ResNet50& ImageNet& Supervised& 25 & 90.22 &  87.62 & 95.15 \\ 
MHIM-MIL \cite{MHIM-MIL} & 20x & ResNet50& ImageNet& Supervised& 25 & 92.48 & 90.75 & 96.49  \\
\hline

HIPT \cite{hipt} & 20x & ViT-s & TCGA & DINO & 22 & NA & NA & 95.7  \\ 
iBot-COAD \cite{iBotCoad} & 20x & ViT-B & TCGA-COAD & iBot & 86 & NA  & NA & 94.5\\ 
CTransPath \cite{CTransPath} & 20x & CNN & TCGA+PAIP& Swin & NA & 92.2 & NA & 94.2\\ 
\Xhline{1pt}
CLAM-SB \cite{CLAM} & 10x & ViT-ti & Cam16 & DINO & 6 &90.69 & 86.36 & 87.60 \\
\rowcolor{green!25}   CLAM-SB \cite{CLAM} & 10x & Vim-ti & Cam16 & DINO & 7 &\textbf{93.02} & \textbf{90.32} & \textbf{95.81} \\
 \hline
\rowcolor{green!25} CLAM-SB\cite{CLAM}& 10x & Vim-ti-plus & Cam16& DINO & 13 & 93.79 & 91.83 & 97.39  \\
 \hline
CLAM-SB \cite{CLAM} & 10x & ViT-s & Cam16& DINO & 22 & \textbf{94.57} & \textbf{92.47} & 96.76  \\
\rowcolor{green!25}  CLAM-SB \cite{CLAM} & 10x & Vim-s & Cam16& DINO & 26 & 92.24 & 89.79 & \textbf{98.85}  \\
\Xhline{1pt}
\end{tabular}
\caption{Results of Slide Level Classification on Camelyon16 \cite{Cam16} dataset. We provide some methods from the literature as a baseline. However, it is essential to note that our method specifically pre-trains on the training set of Camelyon16. As a result, our improvements should not be compared to the other methods in the field. Our goal in this study is to compare Vision Transformers vs Vision Mamba. Furthermore, we acquired ACC and F1-Score values without searching for the optimal threshold. Hence, AUC is the primary metric to compare the performances of these two methods. }
\label{tab:results3}
\end{table*}

\subsection{Experimental setup}
Our experiments were conducted using 4 NVIDIA V100 GPUs, with a batch size of 128 per GPU, achieving an effective batch size of 512. We ensure that this batch size is maintained across our study. We train each model within the DINO framework \cite{DINO} for 100 epochs. The learning rate initiates at 0 and linearly increases over the initial 10 epochs to reach its baseline value of $lr = 0.0005$. Following this initial warm-up, we decay the learning rate using cosine scheduling \cite{sgdr}. We also use weight decay with an initial value of 0.04, which is increased using a cosine scheduler \cite{sgdr} to 0.4. For the second training phase, we freeze each model as the encoder and train a single fully connected layer using cross-entropy loss. We then use the predictions to measure the accuracy of classification performance on the balanced dataset. 

In our experiments, we deploy two versions of Vision Transformer (ViT) encoders: ViT-tiny (ViT-ti) with an embedding dimension of 192 and ViT-Small (ViT-S) with an embedding dimension of 384, both featuring a depth of 12 layers. As highlighted in the Mamba paper \cite{mamba}, a single transformer layer is equivalent to two Mamba block layers. Consequently, for the Vision Mamba tiny (Vim-ti) configuration, we set the depth to 24 layers with an embedding dimension of 192 to maintain comparability. For an enhanced version of Vim-ti, denoted as ViT-ti-plus, we decreased the depth to 12 layers while increasing the embedding dimension to 384. This modification aims to investigate the impact of the embedding dimension size on the performance of Vim models. For the Vision Mamba small (Vim-S) setup, we select an embedding dimension of 384 and a depth of 24 layers. The list of trainable parameters across these configurations is provided in Table \ref{tab:model_parameters}. For slide-level classification, we use the CLAM \cite{CLAM} pipeline to extract patches without using ROI labels. Then, we pass those patches to our encoder model and get the corresponding features. Subsequently, we use the extracted features in the MIL pipelines to classify the whole slides without using the ROI regions. 

\subsection{Results}
In this section, we report the results of our experiments. We first report the performance of both ViT and Vim models for slide-level classification using MIL-based \cite{MIL} methods. Subsequently, we report the performance on patch-level classification on PCam-224 datasets extracted using the ROI labels of Camelyon16.    

\textbf{Slide-Level Classification} For slide-level classification, we use the pre-processing pipeline of CLAM \cite{CLAM} to extract non-background patches from each slide. Then, we use each model as a feature extractor to compute an embedding for that given patch. Subsequently, we use CLAM-SB \cite{CLAM} to classify the slide based on the extracted patch-level feature embeddings. We provide a comparison between our method and other methods on Camelyon16 in Table \ref{tab:results3}. For the baseline results, we acquired them from \cite{MHIM}, \cite{iBotCoad}, and \cite{CTransPath}. We report the F1 Score and Accuracy without searching for an optimal threshold. Hence, while we report them, the ROC AUC is the metric to consider for comparison. The results indicate a substantial improvement in performance with the Vim-ti model, which advances the AUC from 87.60 (as achieved by the ViT-ti model) to 95.81. This increase is particularly significant given the comparable number of parameters involved. This significant enhancement underscores the efficacy of the Vim model in more resource-constrained parameter settings. When considering scaled models, the Vim-ti-plus model, with 13 million parameters, outperforms the 22 million ViT-S model by achieving an AUC of 97.39, compared to the latter's 96.76. This suggests that the Vim model not only competes effectively at larger scales but can also offer efficient improvements. Furthermore, by scaling up to the Vim-s model with 26 million parameters, we see it remains competitive and sets the highest AUC of 98.85 among all models, emphasizing the Vim architecture's consistent performance advantages at various scales. This pattern suggests that Vim models excel compared to ViT models in smaller-scale applications and maintain that level of performance when the number of parameters is increased.

\textbf{Patch-Level Classification} While the main goal of the Camelyon16 dataset is slide-level classification, we also provide the performance of models on the extracted patch-level datasets PCam-224-5x and PCam-224-10x to showcase the learning capability of the proposed method on weakly-supervised classification tasks. For evaluating the patch-level classification performance, we use accuracy for both PCam-224 datasets since they are balanced. The results for PCam-224-10x are provided in Table \ref{tab:results2}. We provide random weights not as a comparison metric but to show the effect of self-supervised learning in the given task. Our observations indicate that the Vim-ti model achieves superior performance over the ViT-ti of a comparable number of parameters. In contrast, ViT-s outperforms the Vim-ti-plus and Vim-S models by a small margin in this task. Furthermore, the performance improvement for the given task from Vim-ti-plus to Vim-S is minimal, while the number of parameters doubles. 

\begin{table}[htp]
    \centering
    \begin{tabular}{c|l|c}
        \Xhline{1.0pt}
        Weights & Model & \begin{tabular}{@{}c@{}}Linear \\ Evaluation  (ACC)\end{tabular} \\ \Xhline{1.0pt}
        \multirow{4}{*}{Random Weights} & ViT-ti & 79.16  \\
        & Vim-ti & 74.87  \\
        & Vim-ti-plus &  77.29\\
        & ViT-s &  83.19\\\hline
        \multirow{4}{*}{Cam16-10x DINO} & ViT-ti & 94.59 \\
        & \cellcolor{green!25}Vim-ti & \cellcolor{green!25}\textbf{95.99} \\
        \cline{2-3}
        & \cellcolor{green!25}Vim-ti-plus & \cellcolor{green!25}96.48 \\ 
        \cline{2-3}
        & ViT-s & \textbf{96.65} \\ 
        & \cellcolor{green!25}Vim-s & \cellcolor{green!25}96.51 \\ \Xhline{1.0pt}
    \end{tabular}
    \caption{Results on PCam-224-10x. }
    \label{tab:results2}
\end{table}

We also perform some tests for 5x zooming level using PCam-224-5x. The results are summarized in the Table \ref{tab:results1}. We provide random weights not as a comparison metric but to show the effect of self-supervised learning in the given task. We have a consistent result as before, with Vim-ti outperforming its alternative ViT-ti with a similar number of parameters. Furthermore, Vim-ti-plus also outperforms ViT-s with approximately half the parameters. 

\begin{table}[htp]
    \centering
    \begin{tabular}{c|l|c}
        \Xhline{1.0pt}
        Weights & Model & \begin{tabular}{@{}c@{}}Linear \\ Evaluation (ACC)\end{tabular} \\ \Xhline{1.0pt}
        \multirow{4}{*}{Random Weights} & ViT-ti & 82.08 \\
        &Vim-ti & 77.12 \\
        & Vim-ti-plus & 78.37 \\
        & ViT-s & 81.61 \\
        \hline
        \multirow{4}{*}[0ex]{Cam16-5x DINO} & ViT-ti & 88.87 \\
        & \cellcolor{green!25}Vim-ti & \cellcolor{green!25} \textbf{90.45} \\
        \cline{2-3}

        & \cellcolor{green!25}Vim-ti-plus & \cellcolor{green!25} \textbf{90.39} \\ 
        & ViT-s & 90.36 \\ \Xhline{1.0pt}
    \end{tabular}
    \caption{Results on PCam-224-5x.}
    \label{tab:results1}
\end{table}
\subsection{Explainability}
We use GradCam \cite{GradCAM} to generate heatmaps of areas that each model is attending to the most by considering the activation maps of the CLS token of each model with respect to the input. Note that while we use the CLS token, both models are trained without labels. 

\begin{figure}[htp]
  \centering
  \includegraphics[width=1\linewidth]{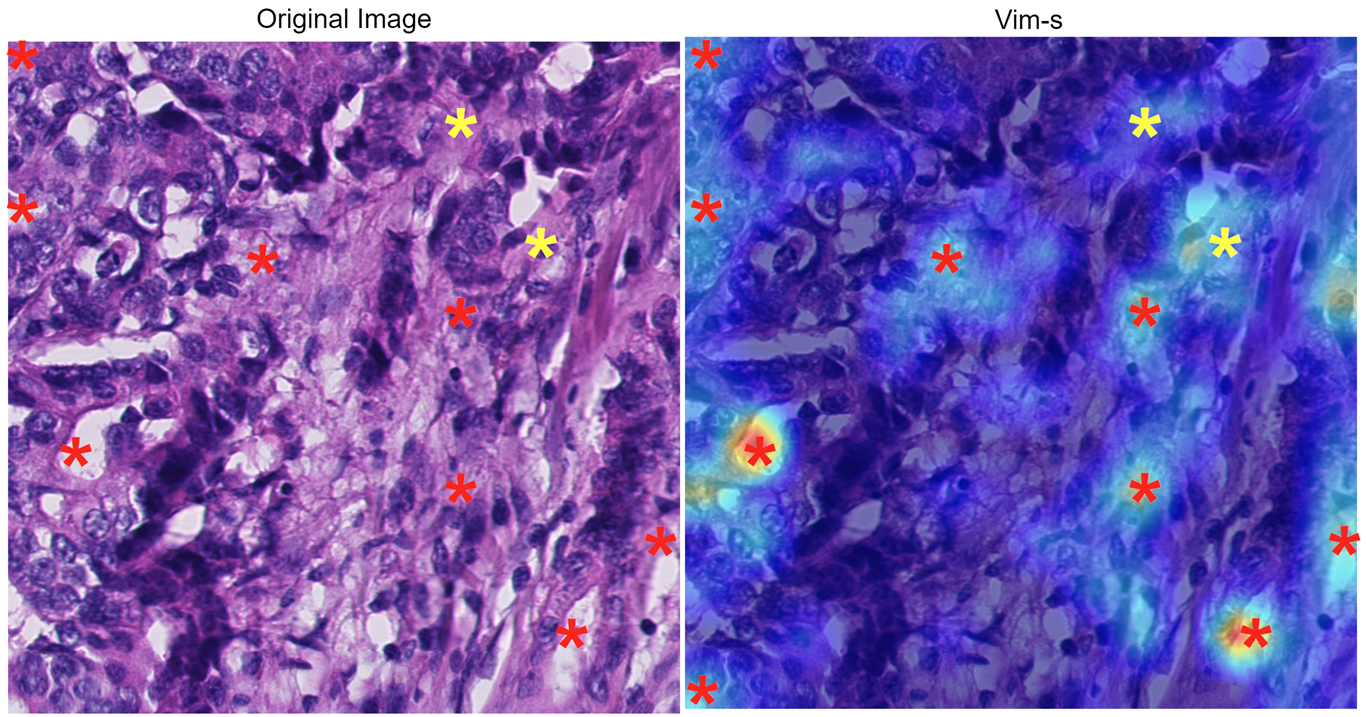}
   \caption{Representative tumor patch with Vim-s heatmap. The red asterisks highlight intracellular mucin in cancer cells. The yellow asterisks highlight stromal features adjacent to cancer cells. (The heatmaps are generated at 10x and overlaid on 40x images.)}
   \label{fig:explain1}
\end{figure}

Our board-certified pathologist reviewed these CLS token activation heat maps generated from each model on a representative subset of tumor patches and compared Vim-based models against ViT-based models. Vim heatmaps were generally oriented foremost to distinctive cancer-specific cellular features and the interface with non-cancer cells. Concurrently, ViT heatmaps foremost highlighted atypical cancer cells. For example, Vim heatmaps highlighted intracellular mucin, a feature nearly 100\% specific for cancer, as well as adjacent activated lymphocytes, which are biologically reactive to the presence of cancer in the lymph node (Fig \ref{fig:explain1}). Indeed, both of these features are exclusively found in the presence of cancer.

\begin{figure}[htp]
  \centering
  \includegraphics[width=1\linewidth]{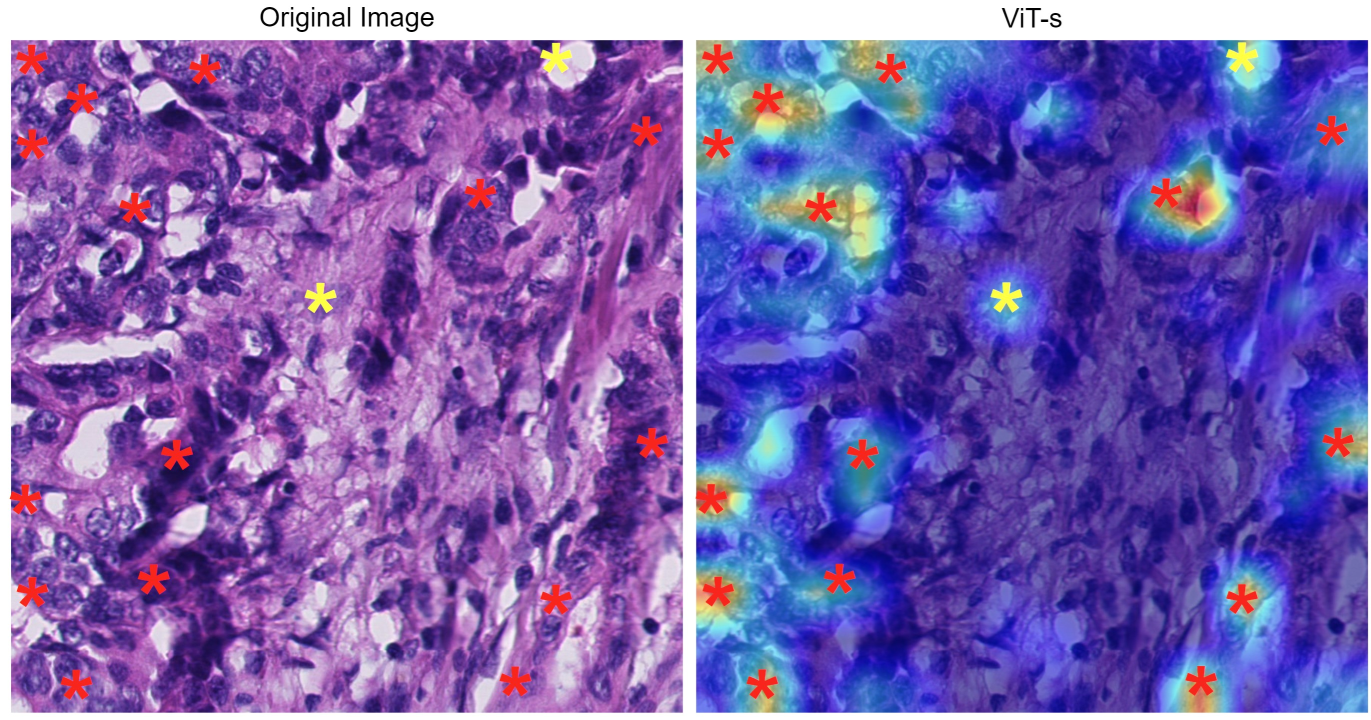}
   \caption{Representative tumor patch with ViT-s heatmap. The red asterisks highlight areas centralized on cancer cells. The yellow asterisks highlight other features, notably a focus of intracellular mucin (top-right) and a stromal cell (middle). (The heatmaps are generated at 10x and overlaid on 40x images.)}
   \label{fig:explain2}
\end{figure}
Alternatively, ViT heatmaps were focused on cancer cells as a whole, notably very atypical cancer cells, nearly in a dichotomous fashion of cancer versus non-cancer (Fig \ref{fig:explain2}). 

Overall, while both models worked efficiently, Vim could identify biological features that pathologists use to validate the presence of cancer inside lymph nodes, while ViT was oriented toward distinguishing whole cancer cells from non-cancer cells. This is an indication that the proposed Vim4Path framework mimics the workflow of pathologists.

\subsection{Ablation Study}
In this section, we provide an ablation study on the impact of zooming level for both slide-level and patch-level classification. Table \ref{tab:results5} shows the impact of transferring a weight trained on one zooming level to another. We notice that doing MIL at the 5x zooming level performs poorly, while MIL at a 10x zooming level gives reasonable results. Furthermore, a model pretrained on 5x does not transfer well to 10x for feature extraction. This performance drop is expected as the data distribution shifts and the fact that the models trained at 5x had fewer pre-training datasets, as summarized in Table \ref{tab:patch_distribution}. The only model that performs reasonably with distribution shift is ViT-s trained on 10x when tested on 5x.     

\begin{table}[htp]
    \centering
    \begin{tabular}{c|l|c|c|c|c}
        \Xhline{1.0pt}
        \begin{tabular}{@{}c@{}}Pretrained \\ Weights \end{tabular} & Model & \multicolumn{2}{c|}{\begin{tabular}{@{}c@{}}MIL for \\   Cam16 at 5x \end{tabular}} & \multicolumn{2}{c}{\begin{tabular}{@{}c@{}}MIL for \\   Cam16 at 10x \end{tabular}} \\
        \cline{3-6} 
         & & AUC & ACC & AUC & ACC \\ \Xhline{1.0pt}
        \multirow{4}{*}{\begin{tabular}{@{}c@{}} Cam16-5x \\ DINO \end{tabular}} & ViT-ti & 65.91 & 73.64 & 68.85 & 65.11 \\
        &\cellcolor{green!25} Vim-ti & \cellcolor{green!25} 71.30 & \cellcolor{green!25} 73.64 &  \cellcolor{green!25} 64.41 & \cellcolor{green!25} 69.76   \\
        & \cellcolor{green!25}Vim-ti* & \cellcolor{green!25}79.03 & \cellcolor{green!25}75.96 & \cellcolor{green!25}74.08 & \cellcolor{green!25}72.86 \\
        & ViT-s & 67.72 & 76.74 & 69.87 & 74.41 \\ \hline
        \multirow{4}{*}{\begin{tabular}{@{}c@{}} Cam16 \\ DINO-10x \end{tabular}} & ViT-ti &  75.05  & 72.86  & 87.60 & 90.69 \\
        & \cellcolor{green!25}Vim-ti &  \cellcolor{green!25}62.14  &  \cellcolor{green!25}54.26 & \cellcolor{green!25}95.81 & \cellcolor{green!25}93.02 \\
        & \cellcolor{green!25}Vim-ti* & \cellcolor{green!25}71.47 & \cellcolor{green!25}79.06 & \cellcolor{green!25}97.39 & \cellcolor{green!25}93.79  \\
        & ViT-s &   84.89 & 81.39 &  96.76 & 94.57  \\ 
        & \cellcolor{green!25}Vim-s &   \cellcolor{green!25}79.64 & \cellcolor{green!25}79.06 &  \cellcolor{green!25}98.85 & \cellcolor{green!25}92.24  \\ \Xhline{1.0pt}
    \end{tabular}
    \caption{Effect of Zooming level on MIL. * stands for Vim-ti-plus}
    \label{tab:results5}
\end{table}

Table \ref{tab:results4} displays the performance of patch-level classification across different zooming levels. It reveals that feature representations learned at one magnification can be applied to other levels in patch classification tasks. Most importantly, models pretrained with a larger dataset at a 10x zooming level (summarized in Table \ref{tab:patch_distribution}) demonstrate superior representation quality during linear evaluation at a 5x level compared to those trained at 5x. This discrepancy shows the significance of dataset size in the pre-training phase in enhancing the quality of the learned representations.

\begin{table}[h]
    \centering
    \begin{tabular}{c|l|c|c}
        \Xhline{1.0pt}
        \begin{tabular}{@{}c@{}}Pretrained \\  Weights \end{tabular}  & Model & \begin{tabular}{@{}c@{}}Linear \\ Eval (5x) \end{tabular} & \begin{tabular}{@{}c@{}}Linear \\ Eval (10x) \end{tabular} \\ \Xhline{1.0pt}
        \multirow{4}{*}{Cam16-5x DINO} & ViT-ti & 88.87 & 89.29 \\
        & \cellcolor{green!25}Vim-ti & \cellcolor{green!25}\textbf{90.45} & \cellcolor{green!25}\textbf{91.03}\\
        & \cellcolor{green!25}Vim-ti* & \cellcolor{green!25}90.39 & \cellcolor{green!25}88.72\\
        & ViT-s & 90.36 & 89.92 \\ \hline
        \multirow{4}{*}{Cam16-10x DINO} & ViT-ti & 90.99 & 94.59 \\
        & \cellcolor{green!25}Vim-ti & \cellcolor{green!25}93.01 & \cellcolor{green!25}95.99 \\
        & \cellcolor{green!25}Vim-ti* & \cellcolor{green!25}\textbf{93.23} & \cellcolor{green!25}96.48 \\
        & ViT-s & 93.00 & \textbf{96.65} \\
        & \cellcolor{green!25}Vim-s & \cellcolor{green!25}93.04 & \cellcolor{green!25}96.51  \\ \Xhline{1.0pt}
    \end{tabular}
    \caption{Effect of Zooming level on patch classification.  * stands for Vim-ti-plus}
    \label{tab:results4}
\end{table}

% To the best of our knowledge, the Vim model \cite{Vim} has not previously been used with the DINO framework \cite{DINO}. Therefore, we have experimented with various learning rate settings for Vim-ti-plus, and the corresponding performance metrics for the PCam-224-5x are displayed in Table \ref{tab:results_lr}. Due to computational constraints, we trained our models for both 5x and 10x magnification levels using the default learning rate settings. As such, refining the hyper-parameters presents a possibility to enhance our results further.

% \begin{table}[htp]
%     \centering
%     \begin{tabular}{l|c}
%         \Xhline{1.0pt}
%         Learning Rate & Linear Evaluation (5x)\\ \hline
%         0.01 & No convergence  \\ \hline
%         0.005 & No convergence  \\ \hline
%         0.001 & 90.16  \\ \hline
%         0.0008 & \textbf{90.95}  \\ \hline
%         0.0007 & 89.71  \\ \hline
%         0.0005 (Default) & 90.39  \\ \hline
%         0.0003 & 89.71  \\ \hline
%         0.0001 & 89.33  \\ \Xhline{1.0pt}
%     \end{tabular}
%     \caption{Learning Rate Ablation study for Vim-ti-plus model}
%     \label{tab:results_lr}
% \end{table}

% \begin{table}[htp]
% \centering
% \begin{tabular}{lccccc}
% \toprule
% Model & TP & FP & TN & FN  & Total\\
% \midrule
% ViT-ti & 38 & 1 & 79 & 11 & 129  \\
% Vim-ti & 42 & 2 & 78 & 7 & 129 \\
% Vim-ti-plus & 45 & 5  & 75 & 4 & 129  \\
% ViT-s  &  43 & 1 & 79 & 6 & 129 \\
% Vim-s  &  44 & 5 & 75 & 5 &  129  \\
% \bottomrule
% \end{tabular}
% \caption{Confusion Matrices for Models using CLAM-SB.}

% \end{table}

\subsection{Discussion}
The Vim model's ability to outperform the ViT at smaller scales is significant. Vim's architecture is potentially more efficient, harnessing fewer parameters to achieve higher accuracy. This efficiency is crucial in scenarios with limited computational resources where cost considerations are essential. Vim is also comparable to ViT on larger scales.  Furthermore, the Vim seems to emulate the pathologists' workflow. These findings suggest that Vim is a promising architecture design for pathology applications that aligns more with clinical workflows. However, our study has limitations. The compute constraints prevented exhaustive hyper-parameter tuning and more extended training regimes, which could unlock even higher performance from the Vim model. Additionally, the performance improvements at different zooming levels raise essential questions about the transferability of features across scales and the role of dataset size in pre-training. Future work should extend the comparative analysis to include more diverse datasets. There is also a need for deeper investigation into the interpretability of the models to build trust in deploying these models in clinical settings.

% In this section, we critically analyze the results obtained from our experiments and situate them within the broader context of computational pathology. Our comparative study between Vision Mamba (Vim) and Vision Transformers (ViT) unveils the nuanced performance differences that are pivotal for practitioners in the field. ===> this is jargon
%The implications for large-scale health data analysis are profound, as it enables more extensive and detailed studies without prohibitive computational costs. ===> this is jargon

% \input{sec/6_pathology}
\section{Conclusion } \label{sec:conclusion}
In this paper, we studied the utility of the Vision Mamba (Vim) encoder model within a Self-Supervised Learning (SSL) framework, particularly DINO, to extract feature embeddings for computational pathology. In particular, we have benchmarked the Camelyon16 dataset to extract image patches from WSIs without labels for training the Vim encoder using the DINO framework. We modified the Vim model to intake arbitrary input image size using positional embedding interpolation, making the Vim model adaptable within DINO for SSL. We compare Vim with ViT in patch-level and slide-level classification tasks on the Camelyon16 dataset, demonstrating Vim's superior performance at small scales. We also show that Vim remains competitive and often outperforms ViT at larger scales. Additionally, our findings reveal that Vim, in contrast to ViT, replicates the process pathologists use to identify cancer. Our study indicates the potential benefits of employing state space models in computational pathology.
{
\small
\bibliographystyle{ieeenat_fullname}
\bibliography{main}
}


\end{document}